# Novel chemistry of lithium oxides and superconducting low-pressure LiO$_4$

Xiao Dong[1,*], Yan-Ling Li[2,*], Artem R. Oganov[3-6], Kuo Li[1,*], Haiyan Zheng[1] and Ho-kwang Mao[1]

[1] Center for High Pressure Science and Technology Advanced Research, Beijing 100193, China

[2] Laboratory for Quantum Design of Functional Materials, School of Physics and Electronic Engineering, Jiangsu Normal University, 221116 Xuzhou, China.

[3] Skolkovo Institute of Science and Technology, Skolkovo Innovation Center, 3 Nobel St., Moscow 143026, Russia

[4] Department of Geosciences and Department of Physics and Astronomy, Stony Brook University, Stony Brook, New York 11794-2100, USA

[5] Moscow Institute of Physics and Technology, 9 Institutskiy Lane, Dolgoprudny city, Moscow Region, 141700, Russia

[6] School of Materials Science, Northwestern Polytechnical University, Xi'an, 710072, China

**We study the stability of Li-O compounds as a function of pressure, with Li ion battery applications and fundamental chemical interest in mind. Using the *ab initio* evolutionary algorithm, we predict stability of novel compounds LiO$_4$, Li$_5$O$_3$ and Li$_6$O under pressure. LiO$_4$, formed at the pressure of just 6 GPa, can be seen as ε-O$_8$ accepting two electrons from two Li atoms. This phase is superconducting, with T$_c$ up to 12.2 K at 10 GPa. This is remarkable, because elemental oxygen becomes superconducting at much higher pressure (96 GPa) and has much lower T$_c$ (<0.6 K), and suggests that chemical alloying with other elements has the potential of not only decreasing metallization pressure, but also of increasing T$_c$. Since ε-O$_8$ is called red oxygen, LiO$_4$ can be identified "lithium red-oxide", and is distinct from superoxide. Additionally, Li$_5$O$_3$ is stable at pressures above 70 GPa and can be represented as a hybrid structure 4Li$_2$O·Li$_2$O$_2$, and electride suboxide Li$_6$O is stable above 62 GPa.**

Correspondence and requests for materials should be addressed to X D.(email:xiao.dong@hpstar.ac.cn), Y. L. L. (email: ylli@jsnu.edu.cn) or K. L. (email:Likuo@hpstar.ac.cn).

Oxygen is one of the most abundant and important elements. However, elemental oxygen is quite unique in many ways: at zero pressure, it is the only known elemental molecular magnet with triplet state as its ground state. The $O_2$ molecule has half-filled $2\pi^*_{p_x}$ and $2\pi^*_{p_y}$ orbitals. At high pressure, the $O_2$ molecules cluster together, forming a non-magnetic ε-phase with $O_8$ cluster [1-3]. It is believed that unpaired bonding orbitals of the neighboring $O_2$ dimers overlap, which induces electron pairing and formation of intermolecular bonds. At higher pressure, metallization is observed around 100 GPa upon the formation of ζ-$O_2$, which is confirmed to have superconductivity at temperatures below 0.6 K [4].

At atmospheric pressure, alkali metal elements are thought to be the most electropositive elements. They adopt the +1 oxidation state and form ionic crystals with electronegative elements. Together with oxygen molecule, they react exothermically and usually form oxides, peroxides or superoxides; heavy alkalis (K, Rb, Cs) are also known to form suboxides (such as $Rb_6O$ and $Rb_9O_2$). Specifically, superoxide anion has one unpaired electron and behaves as a free radical. At low temperatures, superoxides are calculated to be ferromagnetic (FM) or antiferromagnetic (AFM) [5,6] and have some characteristic of d-block and f-block elements: for example, $RbO_2$ is thought to be a Mott insulator[7]. However, compared with d-block elements, the localized unpaired electrons in superoxide have weak exchange effect, so their Curie temperatures are low and are usually paramagnetic state at room temperature. Like most free radicals, such as oxygen molecule, superoxide ion $O_2^-$ has spin pairing and loses its magnetism under pressure. More interestingly, doped with electrons, superoxide ion goes directly to the metallic state, bypassing the stage of semiconducting cluster phase ε-$O_2$.

Lithium is the smallest alkali metal, and its oxide plays an important role in the lithium-air battery. The known oxides are $Li_2O$ and $Li_2O_2$. Some also consider $LiO_2$ to exist in the gas phase[8,9], as the discharge media of Li-$O_2$ battery[10] or on specific substrates, such as graphene[11]. However, pure $LiO_2$ crystals have hardly ever been obtained at normal condition, because $LiO_2$ is thermodynamically unstable with respect to disproportionation, giving $Li_2O_2$[11-13]. Theoretical structure prediction study[14] also reported $Li_3O_4$ to be stable as a hybrid structure of $Li_2O_2 \cdot LiO_2$.

Recent theoretical and experimental investigations found that pressure greatly affects chemical properties of the elements. For example, pressure increases the reactivity of xenon and its oxides become thermodynamically stable at moderate and experimentally reachable pressure (> 83 GPa)[15]. Caesium becomes a *p*-block element and the formation of $CsF_n$ (*n* > 1) compounds was predicted[16,17]. Sodium becomes an extremely electropositive element and forms a very stable compound $Na_2He$ with the normally inert element He at pressures above ~120 GPa[18]. Furthermore, under pressure, unexpected sodium chlorides, such as $Na_3Cl$ and $NaCl_3$, become stable[19].

Since oxygen and superoxide ion lose their unpaired spin characters and their state changes with increasing pressure, we expected that there will be different chemical behaviors in the Li-O system at high pressure. Here we do the structure prediction with a variable-composition evolutionary structure prediction algorithm[20], as implemented in the USPEX code[21]. In such calculations, a phase is deemed stable if its enthalpy of formation from the elements or any other possible compounds is negative. Variable-composition structure searches with spin polarization were performed for the Li-O system at pressures of 0, 10, 20, 50, 70 and 100 GPa, allowing up to 45 atoms per primitive cell. Unexpectedly, we found a series of novel compounds, such as $LiO_4$, $Li_5O_3$, $Li_6O$, which have lower enthalpy than the mixture of elemental Li and O, or any other mixture at pressures below 100 GPa. In our calculations, the ordinary ambient-pressure phases, *Pnnm* $LiO_2$, $P6_3/mmc$ $Li_2O_2$, and *Fm*-3*m* $Li_2O$ are successfully reproduced and their cell parameters are consistent with previous experimental and theoretical works[10,12,14,22-27]. For other theoretically predicted phases, such as *P*-6*m*2 $Li_3O_4$, our calculations (taking great care for the precision and quality of the PAW potential) indicate metastabilty.

$LiO_2$ is found to have a phase transition from *Pnnm* to *P*4/*mbm* at the pressure of 12 GPa (Fig. S1), and this transition is a direct analogue of the NaCl-CsCl structural transition (if one treats the superoxide group $O_2^-$ as a single particle), well known in binary ionic systems at high pressure. In *Pnnm* phase, Li is inside the oxygen layer and has shorter distance (2.35 Å at 10 GPa) to superoxide anion than in *P*4/*mbm* (2.49 Å at 10 GPa), making it easier to bridge the localized spin densities by superexchange interaction, as reported in alkali metal superoxides[5-7]. Thus, this phase has a non-zero magnetic moment in its ground state. When the $O_2^-$ groups are forced closer by pressure, the superexchange interaction competes with spin pairing tendency, which triggers a magnetic transition from high spin (1 $\mu_B$ per $O_2^-$) to low spin (about 0.2 $\mu_B$ per $O_2^-$) at 6 GPa. *Pnnm* phase remains weakly magnetic until the net spin decreases to zero at 35 GPa. In the *P4/mbm* structure, Li and O are in different layers, so Li is far away and unable to bridge the unpaired spin densities of $O_2^-$ groups, and this phase loses its magnetism at a low pressure of 2 GPa. Although devoid of stabilizing superexchange interactions, this phase with 8-fold coordination is denser and the corresponding reduction of the PV-term renders it more stable at high pressure. Thus, in addition to the recent growth of $LiO_2$ on graphene substrate[11], crystalline $LiO_2$ can be made under pressure. Our investigation of thermodynamic stability shows that at atmospheric pressure, $LiO_2$ is metastable, and decomposition into $Li_2O_2+O_2$ (0-6 GPa) or $Li_2O_2+LiO_4$ (6-18 GPa) is favorable. This decomposition at zero pressure is consistent with previous calculations[12,13] and with the known problems of obtaining pure crystalline $LiO_2$. Only at pressures above 18 GPa does $LiO_2$ become stable.

Furthermore, a novel compound $LiO_4$ is calculated to become thermodynamically stable at a remarkably low pressure of 6 GPa, and the following reactions:

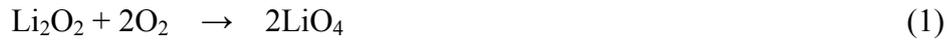
$$Li_2O_2 + 2O_2 \rightarrow 2LiO_4 \qquad (1)$$
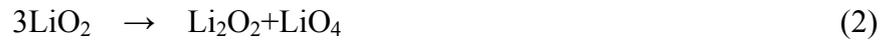
$$3LiO_2 \rightarrow Li_2O_2+LiO_4 \qquad (2)$$

are both predicted to be exothermic at pressures between 6 GPa and 18 GPa, which means $LiO_4$ is predicted to be only stable phase at these pressures in the compositional space from $Li_2O_2$ to pure $O_2$, instead of $LiO_2$. Above 18 GPa, $LiO_2$ and $LiO_4$ coexist up to at least 100 GPa. Phonon calculations (Fig. 2d) clearly indicate dynamical stability of $LiO_4$ from 0 GPa to 100 GPa, which means once formed at high pressure, this novel product can be quenchable to ambient conditions.

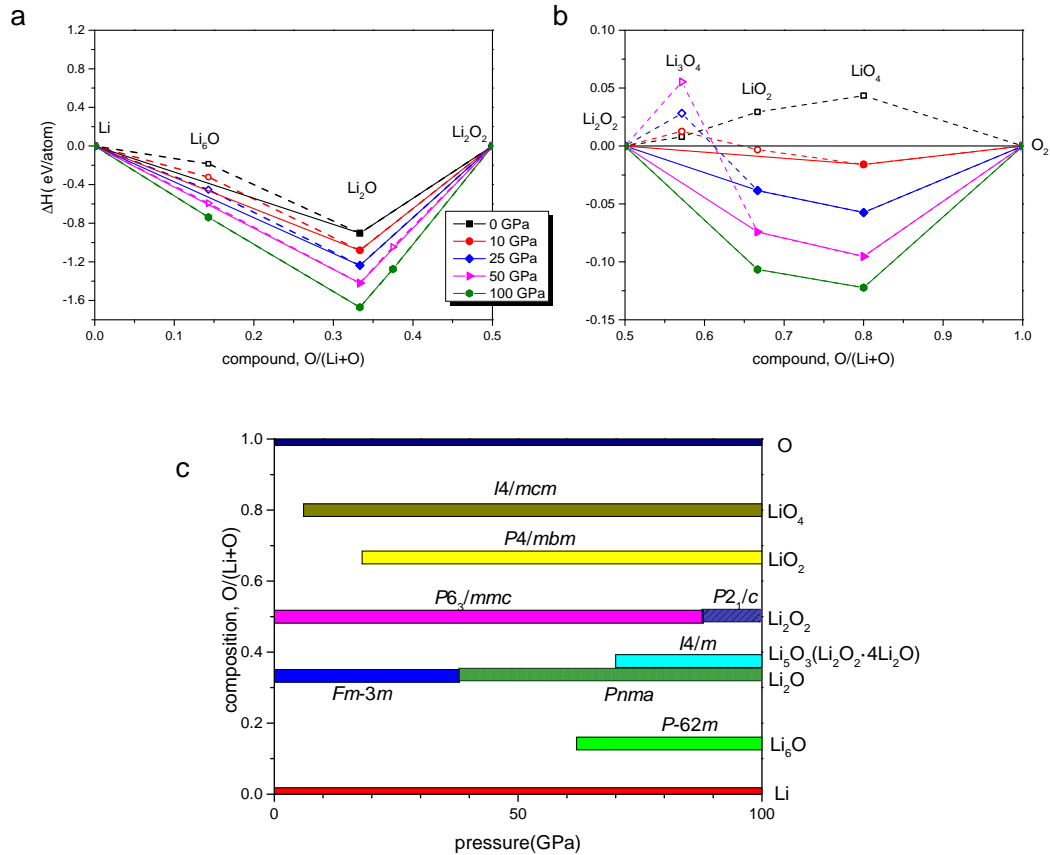

Fig. 1 Thermodynamics of the Li-O system: (a-b) Predicted of convex hulls of the Li-O system at 0, 10, 25, 50 and 100 GPa. For clarity, we show separately (a) Li-rich and (b) O-rich parts of the system. In (c), we show the pressure-composition phase diagram of the Li-O system.

As shown in Fig 1c, $LiO_4$ has only one ground-state structure from 6 GPa to at least 100 GPa, and this structure (shown in Fig. 2) has a body-centered tetragonal space group $I4/mcm$. At 10 GPa, O atoms occupy the Wyckoff position 16l (0.195, 0.305, 0.414), Li atoms occupy the 4a (0, 0, 0.25) positions and the lattice parameters are $a = b = 4.73$ Å, $c = 7.35$ Å. This structure consists of layers of oxygen dimers alternating with Li layers. Particularly, in one conventional cell, oxygen layer is made of 4 oxygen dimers forming an $O_8$ cluster, which looks like a square when viewed along the $O_2$ molecules. The squares in different layers do not align and have a rotation, for example, 25° at 10 GPa. That is why $LiO_4$ is seen like a flower with 8 petals in its [001] direction (Fig. 2b). The O-O distance in $LiO_4$ (1.26 Å at 10 GPa) lies in between that in the neutral oxygen molecule (1.21 Å at 10 GPa) and in the superoxide anion $O_2^-$ (1.32 Å at 10 GPa), which implies an intermediate bonding situation. Band structure shows $LiO_4$ to be a two-dimensional metal, with conductivity along the oxygen layer (Fig. 2c). Fermi level of $LiO_4$ goes through the O-O $\pi^*$ band, like $LiO_2$. Both $LiO_4$ and $LiO_2$ are typical p-type conductors, which can be seen as the saturated system $Li_2O_2$ doped with oxygen dimers. The $\pi^*$-band in *P4/mbm* $LiO_2$ is broader than in $LiO_4$, owing to the crystal geometry and more extra electrons improving the interaction among oxygen -dimers.

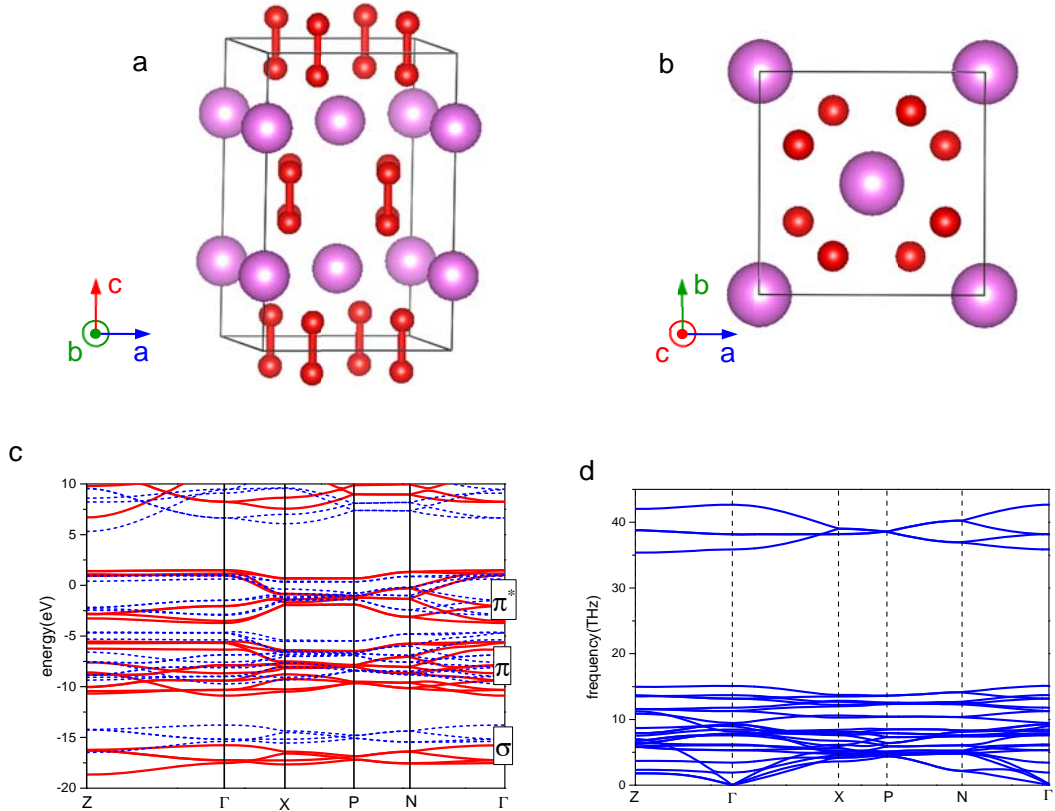

Fig. 2 The conventional cell of $LiO_4$ in (a) [010] and (b) [001] directions. (c) The band structure of $LiO_4$ at 10 GPa. Red solid line are calculated with HSE hybrid functional and blue dotted lines are from the PBE functional. Bands formed by 2p electrons are indicated. Fermi level is set to zero. (d) Phonon spectrum of $LiO_4$ at 0 GPa.

As shown in Fig 4d, the calculated phonon spectrum proves the dynamical stability in the pressure range from 0 to 100 GPa. There is a wide phonon gap up to 20.2 THz at zero pressure in its phonon spectrum to identify $LiO_4$ as a quasi-molecular system with its oxygen dimers. At 10 GPa, frequencies of Raman-active oxygen stretching vibrations are 43.5 ($B_{2g}$) and 37.0 THz ($A_{1g}$), which are intermediate between elemental oxygen $O_2$ (48.0 THz) and *Pnnm* $LiO_2$ (37.0 and 33.1 THz).

The formation of lithium superoxide $LiO_2$ at high pressure is consistent with the existence at ambient pressure of such alkali metal superoxides as $NaO_2$, $KO_2$, $RbO_2$ and $CsO_2$. However, $LiO_4$ was never reported, has no known analogues and is unexpected from classical chemical rules. We can draw, however, some analogies with high-pressure phases $\varepsilon$-$O_8$[1-3] and $NaCl_3$[19]: the former has structural relationship to $LiO_4$ and the latter has electronic similarity with it (it is a metal due to the anion sublattice, just like $LiO_4$). $LiO_4$ can be seen as $\varepsilon$-$O_8$ accepting 2 electrons from Li atoms.

Molecular oxygen, with half-filled $\pi^*$ bands, should be metallic – but opens a band gap through magnetization (Stoner mechanism) and distortion (Peierls mechanism) to lower the energy. In semiconducting $\varepsilon$-$O_8$, there are two kinds of intermolecular distances, 2.12 Å and 2.72 Å at 10 GPa – within and between the $O_8$ clusters. Under pressure, magnetism is suppressed and Peierls distortion vanishes because of its increased elastic energy cost and unfavorable volume increase – and this leads to metallization of oxygen at 96 GPa. Injecting the Peierls-unstable metallic system with electrons can widen its stability field and make it stable at lower pressures – as we see in $NaCl_3$ (Peierls distortion disappears at 48 GPa, compared to ~160 GPa in $Cl_2$) and in $LiO_4$ (metallization and disappearance of the Peierls distortion at just 6 GPa, compared to 96 GPa in pure oxygen). At 10 GPa, the O-O distance between oxygen dimers is 2.42 Å, just the average distance in $\varepsilon$-$O_8$.

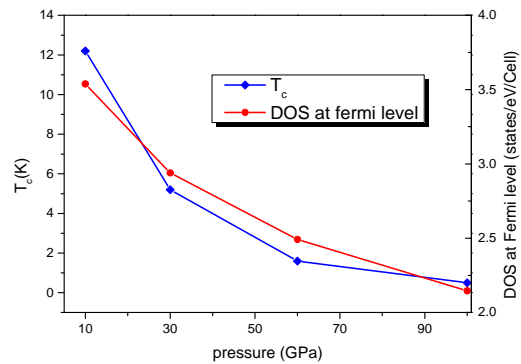

Fig. 3. Critical temperature of superconductivity ($T_c$) and DOS at the Fermi level as a function of pressure.

Inhibition of Peierls transition usually means high electron-phonon coupling,

which leads to superconductivity; for example, ζ-$O_2$ is a superconductor with critical temperature $T_c$ = 0.6 K at 100 GPa[4]. High DOS at the Fermi level and high vibrational frequencies are the other factors raising $T_c$ in Bardeen-Cooper-Schrieffer (BCS) theory[28]. Using Allen-Dynes modified McMillan equation[29] with commonly accepted value of the Coulomb pseudopotential μ* = 0.1, for $LiO_4$ we obtained $T_c$ of up to 12 K at 10 GPa, decreasing with pressure to 0.5 K at 100 GPa (Fig 3), which is quite near the $T_c$ of metallic oxygen molecular at 100 GPa. In case of $LiO_4$, this fall of $T_c$ is due to the sharp decrease of the DOS at Fermi level with pressure (Fig. 3).

Doping by highly electropositive atoms, which donate their electrons to non-metal atoms and can increase the DOS at the Fermi level, can lead to high-$T_c$ superconductors (cf $CaC_6$[30,31]). Here, by adding lithium we not only decrease the pressure of onset of superconductivity to just 6 GPa, but also increase $T_c$ to 12 K (which coincides with $T_c$ of $CaC_6$).

Since ε-$O_8$ is called red oxygen, $LiO_4$ can be called "lithium red-oxide" to emphasize its relationship with ε-$O_8$ and distinguish from superoxide. Metallic and superconducting, $LiO_4$ has a low formation pressure and is predicted to remain dynamically stable at 0 GPa. Since metastable phase $LiO_2$ can grow on specific substrate[11], $LiO_4$ can also be stabilized on properly chosen substrates and might find use in Li batteries.

Upon further increase of pressure, two new compounds become thermodynamically stable: $Li_5O_3$ at 70 GPa and $Li_6O$ at 62 GPa (Fig. 1c). $Li_5O_3$ can be seen as a hybrid structure $4Li_2O \cdot Li_2O_2$, as shown in Fig. 4b; this is a peroxide-oxide compound with simultaneous presence of oxide and peroxide ions at high pressure (such compounds become stable in the Al-O system at pressures above 300 GPa[32]).

$Li_6O$ is a novel kind of suboxide. Like pure Li[33-36], $Li_6O$ at high pressure is an electride with a 1:2 ratio of O ions and localized electron pairs, so the true formula is $Li_6O(2e)_2$. As shown in Fig. 4a, every oxygen and localized electron pair are surrounded by a Li polyhedron of 9 vertexes, 21 edges and 14 triangular faces. Similar electride suboxide, $Mg_3O_2$, was also reported[37] at high pressure. $Li_6O$ is a non-superconducting metal.

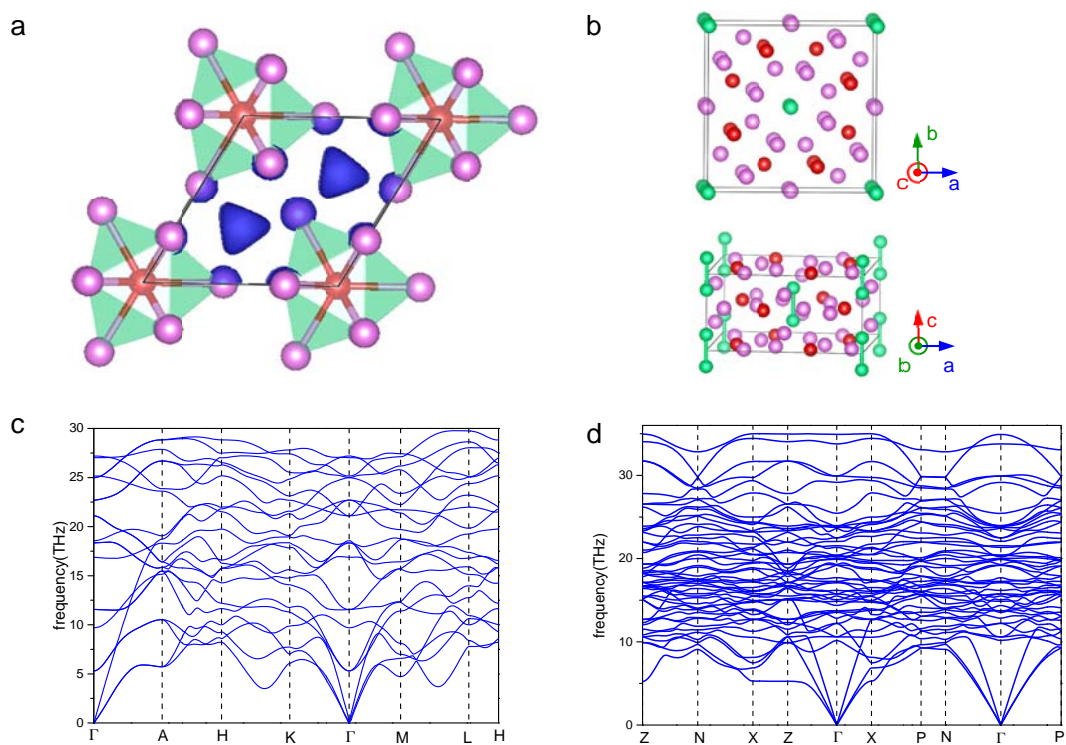

Fig. 4 Crystal structures of (a) $Li_6O$ and (b) $Li_5O_3$ and their phonon spectra at 80 GPa ((c) and (d), respectively). On (a) we also show green isosurface with ELF=0.9. In (b), the structure is shown in [010] and [001] directions with peroxide ions colored green and the oxide ion colored red.

In conclusion, systematic search for stable compounds yielded three novel lithium oxides - $LiO_4$, $Li_5O_3$, $Li_6O$. Of particular interest is $LiO_4$, which becomes stable at just 6 GPa and can be seen as $\varepsilon$-$O_8$ doped with 2 electrons from two lithium atoms. This phase is dynamically stable at atmospheric pressure. With Li injecting its electrons to greatly increase conductivity, $LiO_4$ has electronic conduction within the oxygen layers and superconducting $T_c$ = 12 K, much higher than that of pure oxygen, 0.6 K. After 10 GPa, $T_c$ decreases with the pressure, due to decreasing density of states at the Fermi level for $LiO_4$. These novel predicted compounds not only display unusual chemistry, but also might play a role in Li batteries.

**METHODS**

The evolutionary algorithm USPEX[21], used here for predicting new stable structures, searches for the structure with the lowest enthalpy at specific pressure and is available to predict the stable structure of a compound knowing just the chemical composition.

A number of applications[19,21,38-40] illustrate its power. Structure relaxations were performed using density functional theory (DFT) within the Perdew-Burke-Ernzerhof (PBE) functional[41] in the framework of the all-electron projector augmented wave (PAW) method[42] as implemented in the VASP code[43]. For Li atoms we used PAW potentials with 1.2 a.u. core radius and $1s^22s$ electrons treated as valence; for O the core radius was 1.15 a.u. and $2s^22p^4$ electrons were treated as valence. We used a plane-wave kinetic energy cutoff of 900 eV, and the Brillouin zone was sampled with a resolution of $2\pi \times 0.04$ Å$^{-1}$, which showed excellent convergence of the energy differences, stress tensors and structural parameters. The first generation of structures was created randomly. All structures were relaxed at constant pressure and 0 K, and the enthalpy was used as fitness. The energetically worst structures (40%) were discarded and a new generation was created 30% randomly and 70% from the remaining structures through heredity, lattice mutation and permutation of atoms. For accurate calculations on lowest-enthalpy structures, we used kinetic energy cutoff of 1000 eV and uniform Γ-centered k-meshes with resolution $2\pi \times 0.02$ Å$^{-1}$ for Brillouin zone sampling. Phonon calculations were performed for all promising structures using the PHONOPY code[44]. Electron-phonon coupling and superconductivity calculation was performed in Quantum Espresso package[45].

**ACKNOWLEDGEMENTS**

This work was supported by NSAF (Grant No：U1530402), NSFC (Grant No. 11347007)，Special Program for Applied Research on Super Computation of the NSFC-Guangdong Joint Fund (the second phase), research funds of Skoltech and MIPT, and by the Foreign Talents Introduction and Academic Exchange Program (No. B08040). The calculations were performed at Tianhe II in Guangzhou. We appreciate the discussion with Wenge Yang in High Pressure Synergetic Consortium, Geophysical Laboratory, Carnegie Institution of Washington, USA.

**AUTHOR CONTRIBUTIONS**

X.D., and Y.L.L. designed this research. X.D. and Y.L.L. performed and analyzed the calculations. X.D., A. R.O., and K.L. assisted with calculations. All authors contributed to interpretation and discussion of the data. X.D. and A.R.O wrote the manuscript.